\documentclass[conference]{IEEEtran} % US Letter

\usepackage{etex}

\def\papertitle{Successive-Cancellation Flip Decoding of Polar Codes with a Simplified Restart Mechanism}

\usepackage{ifpdf}
\ifpdf
\usepackage[draft,pdfborder={0 0 0}]{hyperref}

\fi

\usepackage{cite}
\usepackage{graphicx}
\usepackage{epstopdf}

\usepackage{amssymb}
\usepackage[cmex10]{amsmath}
\usepackage[subnum]{cases}
\usepackage{nicefrac}
\usepackage{bm} % bold face for math

\usepackage{algorithm}
\usepackage{algorithmicx}
\usepackage{algpseudocode}

\usepackage{glossaries}
\usepackage{colortbl}
\usepackage{tabularx}
\usepackage{multirow}
\usepackage{dcolumn}
\usepackage{booktabs}

\usepackage{tikz,pgf}
\usepackage{pgfplots}
\pgfplotsset{compat=1.16}
\usetikzlibrary{shapes}
\usetikzlibrary{arrows}
\usetikzlibrary{plotmarks,spy,calc}

\usepackage{balance}
\usepackage{caption}

\RequirePackage{doi}
\usepackage{doi}
\usepackage[markup=underlined]{changes}
\definechangesauthor[color=blue]{PG}
\newcommand{\fixme}[2]{\ifx&#2&{\color{red}#1}\else{\color{red}FIXME\{}#1{\color{red}\}}\footnote{{\color{red}#2}}\PackageWarning{Fixme}{#1: #2}\fi}

\title{\papertitle}

\author{\IEEEauthorblockN{Ilshat Sagitov\IEEEauthorrefmark{1}, Charles Pillet\IEEEauthorrefmark{1}, Alexios Balatsoukas-Stimming\IEEEauthorrefmark{2}, and Pascal Giard\IEEEauthorrefmark{1}}
  \IEEEauthorblockA{\IEEEauthorrefmark{1}Department of Electrical Engineering, \'Ecole de technologie sup\'erieure, Montr\'eal, Qu\'ebec, Canada.\\Email: \{ilshat.sagitov.1, charles.pillet.1\}@ens.etsmtl.ca,  pascal.giard@etsmtl.ca}%
  \IEEEauthorblockA{\IEEEauthorrefmark{2}Department of Electrical Engineering, Eindhoven University of Technology, Eindhoven, The Netherlands.\\Email: a.k.balatsoukas.stimming@tue.nl}
}

\algnewcommand{\Inputs}[1]{%
  \State \textbf{Inputs:}
  \Statex \hspace*{\algorithmicindent}\parbox[t]{.8\linewidth}{\raggedright #1}
}

\algnewcommand{\Initialize}[1]{%
  \State \textbf{Initialize:}
  \Statex \hspace*{\algorithmicindent}\parbox[t]{.8\linewidth}{\raggedright #1}
}

\definecolor{MyBlue}{rgb}{0.00000,0.44700,0.74100}%
\definecolor{MyDarkGreen}{HTML}{006100}
\definecolor{MyMagenta}{rgb}{0.6, 0.4, 0.8}
\definecolor{MyOrange}{rgb}{0.93, 0.57, 0.13}
\newcommand{\plotfigureheight}{0.62}

\makeglossaries

\begin{document}

\bstctlcite{IEEEexample:BSTcontrol}

%% Definition of acronyms
\newacronym{awgn}{AWGN}{additive white Gaussian noise}
\newacronym{snr}{SNR}{signal-to-noise ratio}
\newacronym{fer}{FER}{frame-error rate}
\newacronym{sc}{SC}{successive-cancellation}
\newacronym{scl}{SCL}{successive-cancellation list}
\newacronym{scf}{SCF}{successive-cancellation flip}
\newacronym{dscf}{DSCF}{dynamic \gls{scf}}
\newacronym{crc}{CRC}{cyclic-redundancy check}
\newacronym{llr}{LLR}{log-likelihood ratio}
\newacronym{cc}{CC}{clock cycle}
\newacronym{srm}{SRM}{simplified restart mechanism}
\newacronym{rhs}{RHS}{right-hand side}
\newacronym{lhs}{LHS}{left-hand side}

\maketitle

\begin{abstract}
Polar codes are a class of error-correcting codes that provably achieve the capacity of practical channels. The \gls{scf} decoder is a low-complexity decoder that was proposed to improve the performance of the \gls{sc} decoder as an alternative to the high-complexity \gls{scl} decoder. 
The \gls{scf} decoder improves the error-correction performance of the \gls{sc} decoder, but the variable execution time and the high worst-case execution time pose a challenge for the realization of receivers with fixed-time algorithms. The \gls{dscf} variation of the \gls{scf} decoder further improves the error-correction performance but the challenge of decoding delay remains. In this work, we propose a \gls{srm} that reduces the execution time of \gls{scf} and \gls{dscf} decoders through conditional restart of the additional trials from the second half of the codeword. We show that the proposed mechanism is able to improve the execution time characteristics of \gls{scf} and \gls{dscf} decoders while providing identical error-correction performance. For a \gls{dscf} decoder that can flip up to 3 simultaneous bits per decoding trial, the average execution time, the average additional execution time and the execution-time variance are reduced by approximately 31\%, 37\% and 57\%, respectively. For this setup, the mechanism requires approximately 3.9\% additional memory.

\end{abstract}

\glsresetall

\section{Introduction}
\label{sec:intro}
Polar codes~\cite{arik_polariz} are a type of linear error-correction codes which can achieve the channel capacity for practically relevant channels under low-complexity \gls{sc} decoding. However, at short to moderate block lengths, the \gls{sc} algorithm provides an error-correction performance that is lacking for many practical applications. To address this, the \gls{scl} decoding algorithm was proposed~\cite{scl_intro}. It provides great error-correction capability to the extent that polar codes were selected to protect the control channel in 3GPP's next-generation mobile-communication standard (5G), where \gls{scl} serves as the error-correction performance baseline~\cite{3GPP_5G_Coding}.
However, the error-correction capability of the \gls{scl} decoder comes at the cost of high hardware implementation complexity and low energy efficiency~\cite{scl_5g}.

As an alternative to \gls{scl} decoding, the \gls{scf} decoding algorithm was proposed~\cite{scf_intro}.
\Gls{scf} leads to an improved error-correction performance compared to \gls{sc}, but still falls behind the \gls{scl} decoder with a moderate list size. However, the \gls{scf} decoder is more efficient than \gls{scl} both in terms of computing resources and energy requirements~\cite{Giard_JETCAS_2017}.
\Gls{dscf} decoding, proposed in~\cite{dyn_scf}, significantly improves the error-correction performance of \gls{scf} decoding. 
\Gls{dscf} implements a better metric to identify bit-flipping candidates and the multiple bit-flipping methodology. Preliminary results from a hardware implementation indicate that \gls{dscf} decoders have a higher energy efficiency compared to \gls{scl} decoders with moderate list sizes while providing similar error-correction performance~\cite{pract_dscf}.

Both \gls{scf} and \gls{dscf} decoders exhibit a variable execution time and the variance of that execution time can be significant. This poses a challenge in the realization of receivers, where fixed-time algorithms are preferred. An early-stopping mechanism for the single bit-flip \gls{dscf} decoder that aims to reduce the execution-time characteristics was proposed in~\cite{earl_stop_dscf}. However, it negatively affects the error-correction performance.

\subsubsection*{Contributions}
In this work, we propose a \gls{srm} that reduces the average execution time, the average additional execution time and the execution-time variance of \gls{scf} and \gls{dscf} decoders for polar codes. The central idea of the mechanism is to conditionally restart additional decoding trials from the second half of the codeword by using computations stored following the initial \gls{sc} pass. The error-correction performance is identical to the original non-\gls{srm} decoders. For the multi bit-flip version of \gls{dscf}, the average execution time, the average additional execution time and the execution-time variance are reduced by $31-58\%$, while the additional memory overhead is $2.9-3.9\%$.

\subsubsection*{Outline} 

The remainder of this paper is organized as follows. \autoref{sec:backgr} provides an introduction to polar codes, briefly describes \gls{sc}, \gls{scf} and \gls{dscf} decoders. In \autoref{sec:simp_trial}, the \gls{srm} is presented, where an algorithm is described along with memory requirements. In \autoref{sec:lat_hw_scf}  execution-time characteristics of decoders under hardware constraints are discussed. In \autoref{sec:sim_res}, the simulation methodology and results are presented. \autoref{sec:conclusion} concludes the work.

\section{Background}
\label{sec:backgr}
\subsection{Construction of Polar Codes}
A polar code $\mathcal{P}(N,k)$, where $N=2^n$ is the code length and $k$ is the code dimension, relies on the channel polarization induced by $\bm{G}^{\otimes n}$, defined as the $n^{\text{th}}$ Kronecker power of the binary kernel $\bm{G}=\left[ \begin{smallmatrix} 1 & 0 \\ 1 & 1\end{smallmatrix} \right]$. The $\left(N-k\right)$ least-reliable bits, called frozen bits, are set to predefined values that are known by the decoder, typically all zeros. The $k$ information bits are set to the most reliable positions and the code rate is $R=\nicefrac{k}{N}$.
The encoding is performed as $\bm{x}=\bm{u} \bm{G}^{\otimes n}$, where $\bm{x}$ and $\bm{u}$ are a codeword and an input vector, respectively. The input vector $\bm{u}$ contains the $k$ information bits in their predefined locations as well as the frozen-bit values. We denote the set of frozen bit indices of the input vector by $\mathcal{A}^C$ and the set of information bit indices by $\mathcal{A}$.  The bit-location reliabilities depend on the channel type and conditions. In this work, the \gls{awgn} channel is considered and the construction method used is that of Tal and Vardy~\cite{tal_constr}. 

\subsection{Successive-Cancellation Decoding}
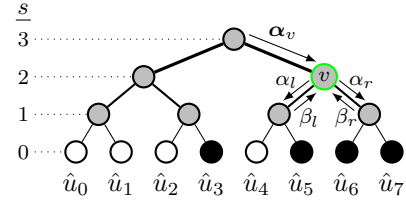
\begin{figure}[t]
	\centering
	\begin{tikzpicture}[]

\node [draw, circle, line width=0.8pt, inner sep=0.1 cm, label=south:{$\hat{u}_0$}] (u0) at (0,0){};
\node [draw, circle, line width=0.8pt, inner sep=0.1 cm, label=south:{$\hat{u}_1$}] (u1) at (0.6,0){};
\node [draw, circle, line width=0.8pt, inner sep=0.1 cm, label=south:{$\hat{u}_2$}] (u2) at (1.2,0){};
\node [draw, circle, line width=0.8pt, inner sep=0.1 cm, label=south:{$\hat{u}_3$}, fill=black] (u3) at (1.8,0){};
\node [draw, circle, line width=0.8pt, inner sep=0.1 cm, label=south:{$\hat{u}_4$}] (u4) at (2.4,0){};
\node [draw, circle, line width=0.8pt, inner sep=0.1 cm, label=south:{$\hat{u}_5$}, fill=black] (u5) at (3.0,0){};
\node [draw, circle, line width=0.8pt, inner sep=0.1 cm, label=south:{$\hat{u}_6$}, fill=black] (u6) at (3.6,0){};
\node [draw, circle, line width=0.8pt, inner sep=0.1 cm, label=south:{$\hat{u}_7$}, fill=black] (u7) at (4.2,0){};

\node [draw, circle, line width=0.8pt, inner sep=0.1 cm, fill=gray!50!white] (n10) at (0.3,0.5){};
\node [draw, circle, line width=0.8pt, inner sep=0.1 cm, fill=gray!50!white] (n11) at (1.5,0.5){};
\node [draw, circle, line width=0.8pt, inner sep=0.1 cm, fill=gray!50!white] (n12) at (2.7,0.5){};
\node [draw, circle, line width=0.8pt, inner sep=0.1 cm, fill=gray!50!white] (n13) at (3.9,0.5){};
\node [draw, circle, line width=0.8pt, inner sep=0.1 cm, fill=gray!50!white] (n20) at (0.9,1.0){};
\node [draw, circle, color=green,  line width=0.8pt, inner sep=0.05 cm, fill=gray!50!white] (n21) at (3.3,1.0){\textcolor{black}{\footnotesize{$v$}}};
\node [draw, circle, line width=0.8pt, inner sep=0.1 cm, fill=gray!50!white] (n30) at (2.1,1.5){};

\node[inner sep=0] (s00) at (-0.7, 1.9) {\small{$s$}};
\draw ($(s00.south)+(-0.1,-0.07)$)--($(s00.south)+(0.1,-0.07)$);

\node[inner sep=2] (s3) at (-0.7, 1.5) {\footnotesize{$3$}};
\node[inner sep=2] (s2) at (-0.7, 1.0) {\footnotesize{$2$}};
\node[inner sep=2] (s1) at (-0.7, 0.5) {\footnotesize{$1$}};
\node[inner sep=2] (s0) at (-0.7, 0.00) {\footnotesize{$0$}};
 \draw[dotted] (s3)--(n30);
 \draw[dotted] (s2)--(n20);
 \draw[dotted] (s1)--(n10);
 \draw[dotted] (s0)--(u0);
\draw (u0) -- node{}(n10);
\draw (u1) -- node{}(n10);
\draw (u2) -- node{}(n11);
\draw (u3) -- node{}(n11);
\draw (u4) -- node{}(n12);
\draw (u5) -- node{}(n12);
\draw (u6) -- node{}(n13);
\draw (u7) -- node{}(n13);
\draw [line width=0.8pt] (n10) -- node{}(n20);
\draw [line width=0.8pt] (n11) -- node{}(n20);
\draw [line width=0.8pt] (n12) -- node{}(n21);
\draw [line width=0.8pt] (n13) -- node{}(n21);
\draw [line width=1.2pt] (n20) -- node{}(n30);
\draw [line width=1.2pt] (n21) -- node{}(n30);
\draw [-latex] (2.3,1.56) -- node[anchor=south] {\footnotesize{$\bm{\alpha}_{v}$}}(3.2,1.2);
%\draw [dashed, -latex] (3.0,1.0) -- node[anchor=north east] {\footnotesize{$\beta_v$}}(2.2,1.33);

\draw [-latex] (3.1,0.95) -- node{} (2.75,0.68);
\draw [-latex] (2.9,0.55) -- node{} (3.2,0.79);
\draw [-latex] (3.5,0.95) -- node{} (3.85,0.68);
\draw [-latex] (3.7,0.55) -- node{} (3.4,0.79);

\node [] (al) at (2.8,0.93){\footnotesize{$\alpha_l$}};
\node [] (ar) at (3.8,0.93){\footnotesize{$\alpha_r$}};
\node [] (bl) at (3.1,0.45){\footnotesize{$\beta_l$}};
\node [] (br) at (3.6,0.45){\footnotesize{$\beta_r$}};

\end{tikzpicture}
	\caption{\gls{sc} decoding tree of $\mathcal{P}\left(8,4\right)$ polar code.}
	\label{fig:pol_tree}
\end{figure}

The \gls{sc} decoding schedule can be represented as a binary tree traversal through the layers $s\in \{0,\ldots, n \}$ starting from the root node ($s=n$) with the message passing to the \gls{lhs} and then to the \gls{rhs} of the decoding tree. The decoding tree of a $\mathcal{P}\left(8,4\right)$ polar code is shown in \autoref{fig:pol_tree}. 
The received vector of channel \glspl{llr}, denoted by $\bm{\alpha}_{\text{ch}} = \left[\alpha_{\text{ch}}(0),\ldots, \alpha_{\text{ch}}(N-1)\right]$, is at the tree root. 
The $v^{th}$ intermediate node, located in layer $s$, having input vector $\bm{\alpha}_v\in\mathbb{R}^{2^s}$ forwards message $\bm{\alpha}_l\in\mathbb{R}^{2^{s-1}}$ to its left $l$ and $\bm{\alpha}_r\in\mathbb{R}^{2^{s-1}}$ to its right $r$ as:
\begin{align}
    \alpha_l(j)&=f\left(\alpha_v(j),\alpha_v\left(j+2^{s-1}\right)\right),\label{eq:f}\\
    \alpha_r(j)&=g\left(\alpha_v(j),\alpha_v\left(j+2^{s-1}\right),\beta_l(j)\right)\label{eq:g},
\end{align}
with $0\leq j<2^{s-1}$ and  the $f:\mathbb{R}^2\rightarrow\mathbb{R}$ function is the boxplus operator whose hardware-friendly implementation is:
\begin{align}
    f(a,d)=\mathrm{sign}\left(a \right) \cdot \mathrm{sign} \left( d \right) \cdot \mathrm{min}\left( |a|,|d| \right)
\end{align}
and the $g:\mathbb{R}^2\times\mathbb{F}_2\rightarrow\mathbb{R}$ function is defined as:
\begin{align}
    g(a,d,b)=\left( 1-2b \right) \cdot a + d\,.
\end{align}

\glspl{llr} at the leaf nodes of the tree are called decision \glspl{llr} and denoted by $\bm{\alpha}_{\text{dec}} = \left[\alpha_{\text{dec}}(0),\ldots, \alpha_{\text{dec}}(N-1)\right]$. Each information bit from the transmitted vector $\bm{\hat{u}}=\left[\hat{u}_0,\ldots, \hat{u}_{N-1}\right]$ is estimated by taking a hard decision on the corresponding decision \gls{llr}. Frozen bits are known to the decoder and thus directly estimated. Nodes of decision \glspl{llr} corresponding to information bits are in black and of frozen bits are in white in \autoref{fig:pol_tree}. 
Bit-estimates are propagated from lower to higher layers of the tree and used for calculations of partial-sums.  
The vector of partial-sums, denoted by $\bm{\beta}$, is calculated for node $v$ at layer $s$ (\autoref{fig:pol_tree}) as follows:
\begin{equation}
\beta_v\left(j\right) = 
\begin{cases}
\beta_l\left(j\right) \oplus \beta_r\left(j\right) ~~~ \mathrm{if} ~~ j < 2^{s-1} \,,\\
\beta_r\left(j\right) \quad \quad \quad ~~~\;\, \mathrm{otherwise}\,,
\end{cases}
\label{eq:ps_comp}
\end{equation}
where operator $\oplus$ is bitwise XOR operation.

\subsection{SC-Flip Decoding}
\label{sec:intro_scf}
The \gls{scf} decoding algorithm is introduced in~\cite{scf_intro}, where the authors observed that if the first erroneously-estimated bit could be detected and corrected before resuming \gls{sc} decoding, the error-correction capability of the decoder would be greatly improved. In order to detect decoding failure of the codeword, information bits are concatenated with a $r$-bit \gls{crc} being passed through the polar encoder. The \gls{crc} bits extend the set of information bits $\mathcal{A}$ of the polar code,  increasing the code rate to $R=\nicefrac{(k+r)}{N}$. %We further denote the polar code by $\mathcal{P}(N,k+r)$ accordingly.

If decoding failure is identified at the end of the initial \gls{sc} decoding pass, a list of bit-flipping candidates, denoted by $\bm{\mathcal{B}}_{\text{flip}}$, is constructed. The information bit indices with the smallest metrics are identified with the absolute values of $\bm{\alpha}_{\text{dec}}$ being the metrics. The bit-flipping indices are stored in  $\bm{\mathcal{B}}_{\text{flip}}$ in ascending order of their corresponding metrics.

In order to constrain the decoding delay of \gls{scf} decoding, the maximum number of trials $T_{\text{max}}$ is defined, where $T_{\text{max}} \in \mathbb{N}^+$ and $1\leq T_{\text{max}} \leq \left( k+r+1\right)$, including the initial \gls{sc} pass. When additional trials are performed, one bit from $\bm{\mathcal{B}}_{\text{flip}}$ is selected for flipping. Setting $T_{\text{max}}=1$ renders \gls{scf} equivalent to \gls{sc} decoding. If the \gls{crc} fails after $T_{\text{max}}$ trials, the decoding is stopped and failure is declared. We highlight that $T_{\text{max}}-1$ is the total number of flipping candidates.

\subsection{Dynamic SC-Flip Decoding}
\Gls{dscf} decoding is proposed in~\cite{dyn_scf} with two major improvements to original \gls{scf}. First, a more accurate metric for constructing $\bm{\mathcal{B}}_{\text{flip}}$ is derived. Second, a methodology of flipping multiple bits is proposed, i.e., the decoder is able to flip more than one bit per decoding trial. 

Flipping of multiple bits is achieved by progressively updating the set of bit-flipping candidates $\bm{\varepsilon}_t=\{i_{\lambda}\}$, where $t$ is index of additional trial. The current set size is denoted by $\lambda$,  $1 \leq \lambda \leq \omega$ and  $i_1\leq i_{\lambda} \leq i_{\omega}$. The maximum set size, or decoding order~\cite{dyn_scf}, is denoted by $\omega$ and  indicates the maximum number of bit-flips per trial.

For each bit-flipping set $\bm{\varepsilon}_t$ the metric calculation and update are performed according to:
\begin{equation}
\mathcal{M}_{\text{flip}}(\bm{\varepsilon}_t) = \sum_{j\in \bm{\varepsilon}_t} |\alpha_{\text{dec}}(j)| + \Pi\left(\bm{\varepsilon}_t\right)\,, 
\label{eq:metr_dscf_w}
\end{equation}
where $\Pi(\bm{\varepsilon}_t)$ is:
\begin{equation}
\Pi\left(\bm{\varepsilon}_t\right)=\frac{1}{c} \sum_{\substack{j \leq i_{\lambda} \\ j\in \bm{\mathcal{A}}}} \ln\left(1+e^{-c\cdot |\alpha_{\text{dec}}(j)|}\right) \,,
\label{eq:metr_log_part}
\end{equation}
 where $0 < c \leq 1$.
The value of $c$ is optimized for different block lengths, rates and channel conditions. Similarly to \gls{scf} decoding, each set $\bm{\varepsilon}_t$ is stored in list $\bm{\mathcal{B}}_{\text{flip}}$ in ascending order of metric  $\mathcal{M}_{\text{flip}}\left(\bm{\varepsilon}_t\right)$\,.

After an initial \gls{sc} pass with a decoding failure, the bit-flipping candidates are constructed similarly to \gls{scf}, but with the metric \eqref{eq:metr_log_part} with $\lambda=1$. If the maximum set size is $\omega=1$, no additional metric updates are performed: bit-flipping candidates are sorted in ascending order of metrics and each bit flip is chosen accordingly for additional decoding trial. If $\omega>1$, multiple bit flips are applied and each decoding set is updated at every unsuccessful decoding attempt. At each attempt, a new information index is progressively inserted to a temporary constructed set and the metric update is performed for this set. If metric of the temporary set exceeds the largest metric of the list, it is discarded. If not, it is added to the list while keeping the metric list sorted. The set is not extended further after reaching the maximum size $\omega$.

The decoder is called \gls{dscf}-$\omega$ to emphasize the dependence on the parameter $\omega$. A total of $T_{\text{max}}$ trials are run with a total of $T_{\text{max}}-1$ bit-flipping sets. Thus, the index of sets is in the range of $1 \leq t\leq T_{\text{max}}-1$. We highlight that $T_{\text{max}}\geq (k+r+1)$ is applicable when $\omega>1$, since multiple sets resulting from one single or multi bit-flipping set can be constructed and used as the bit-flipping candidates. 

\subsubsection*{Metric Approximation}
The metric $\Pi$ \eqref{eq:metr_log_part} contains logarithmic and exponential computations. To make the metric updates more hardware-friendly, an approximation is proposed in~\cite{simp_dscf}. We denote it as $\Pi'$ and it is defined as:
\begin{equation}
\Pi'(\bm{\varepsilon}_t) = \begin{cases}
1.5\,, \quad \text{if} \;\; |\alpha_{\text{dec}}(j)|\leq 5.0\,, \\
0\,, \quad \;\;\,\text{otherwise}.
\end{cases}
\label{eq:metr_dscf_w_simp}
\end{equation}
This approximation was shown to result in a negligible coding loss~\cite{simp_dscf,pract_dscf}.
In the remainder, the approximation $\Pi'$ is used for the metric calculations and update of \gls{dscf}-$\omega$ decoding.

\section{Simplified Restart Mechanism}
\label{sec:simp_trial}
In this section, we describe our proposed simplified restart mechanism (SRM) for \gls{scf} and \gls{dscf}-$\omega$ decoding.
The \gls{srm} conditionally avoids redundant computations by storing the necessary bits obtained during the first \gls{sc} pass into an additional memory. This section also provides a memory analysis.

\subsection{Description of the SRM}\label{subsec:SRM_description}
Each additional trial in \gls{scf} decoding (and its variants) starts by redoing the \gls{sc} computations to estimate the very first information bit, and then proceeds all the way to the location that corresponds to the information bit that needs to be flipped. However, we observe that decoding of both bits $\hat{u}_0$ and $\hat{u}_{\nicefrac{N}{2}}$ begins from the root layer $s=n$ of the decoding tree, where channel \glspl{llr} are used. The latter are constant throughout \gls{scf} decoding of the current codeword. Therefore, if the information bit that needs to be flipped is located on the \gls{rhs} tree, intermediate \gls{llr} calculations of the \gls{lhs} tree can be entirely avoided. These observations are independent from the specific patterns of information and frozen bits. The estimated bits and partial-sum results of the \gls{lhs} tree are still required for the \gls{rhs} \gls{sc} computations. 

Due to channel polarization, the information bits are predominantly located at the \gls{rhs} of the decoding tree. In \autoref{fig:pol_tree}, $\nicefrac{3}{4}$ of the information bits are on the \gls{rhs}. Naturally, bit flips in \gls{scf} will often occur on the \gls{rhs}. 
For each additional trial where the flipping index is on the \gls{rhs}, we propose to skip the (unchanged) \gls{lhs} of the decoding tree, i.e., keep the initial $\left[\hat{u}_0,\ldots,\hat{u}_{\nicefrac{N}{2}-1}\right]$ and decode $\left[\hat{u}_{\nicefrac{N}{2}},\ldots,\hat{u}_{N-1}\right]$. To do so, the \gls{lhs} computations from the initial \gls{sc} pass must be available, i.e., the partial sums $\bm{\beta}_{\text{rest}}$ and estimated bits $\bm{\hat{u}}_{\text{rest}}$.
\begin{algorithm}
\footnotesize%\small
\caption{\gls{scf} decoding embedding the \gls{srm}.}
\label{alg:scf_srm}
\begin{algorithmic}[1]
\Procedure{SCF\_ With\_SRM}{$\bm{\alpha}_{\text{ch}}, \bm{\mathcal{A}},  T_{\text{max}}$}
\State $\bm{\beta}_{\text{rest}} \gets \left[0,0,\ldots, 0\right]$
\State $\bm{\hat{u}}_{\text{rest}} \gets \left[0,0,\ldots, 0\right]$
\State $\psi_{\text{rest}}\gets \nicefrac{N}{2}-1$
\For{$t=1, t\leq T_{\text{max}}, t=t+1$}
\If{$t>1$}\color{blue}\Comment{Only at additional trials}
\color{black}
\If{$\bm{\mathcal{B}}_{\text{flip}}(t)>\psi_{\text{rest}}$} 
\color{blue}\Comment{Activate \gls{srm} if needed}
\color{black}
\State $\bm{\beta}\left(0,\ldots,\psi_{\text{rest}}\right) \gets \bm{\beta}_{\text{rest}}$ 
\color{blue}\Comment{Load partial sums}
\color{black}
\State $\bm{\hat{u}}\left(0,\ldots,\psi_{\text{rest}}\right) \gets \bm{\hat{u}}_{\text{rest}}$ 
\color{blue}\Comment{Load bit estimates}
\color{black}
%\State $i_{rest}\gets \nicefrac{N}{2}+1$
\State $srm\_act \gets True$ 
\color{blue}\Comment{Set \gls{srm} flag}
\color{black}
\Else \State $srm\_act \gets False$ 
\color{blue}\Comment{Reset \gls{srm} flag}
\color{black}
\EndIf
\Else \State $srm\_act \gets False$ 
\EndIf
\State$\left(\bm{\hat{u}}, \bm{\alpha_{\text{dec}}}\right) \gets \mathrm{\gls{sc}}\left(\bm{\alpha}_{\text{ch}}, \bm{\mathcal{A}}, \mathcal{B}_{\text{flip}}(t), \bm{\beta}, srm\_act \right)$
\If{$\gls{crc}\left(\bm{\hat{u}} \right)=failure$}
\If{$t=1$}
\color{blue}\Comment{Only at initial trial}
\color{black}
\State  $\bm{\mathcal{B}}_{\text{flip}} \gets   \mathrm{Init\_Flip\_Set}\left(\bm{\alpha}_{\text{dec}}, T_{\text{max}}-1 \right)$
\State $\bm{\beta}_{\text{rest}} \gets \bm{\beta}\left(0,\ldots,\psi_{\text{rest}} \right)$
\color{blue}\Comment{Store partial sums}
\color{black}
\State $\bm{\hat{u}}_{\text{rest}} \gets \bm{\hat{u}}\left(0,\ldots,\psi_{\text{rest}} \right)$ 
\color{blue}\Comment{Store bit estimates}
\color{black}
\Else \State continue
\EndIf
\Else \State break
\EndIf
\EndFor
\State \textbf{return} {$\bm{\hat{u}}$}
\EndProcedure
\end{algorithmic}
\end{algorithm}

Algorithm\,\ref{alg:scf_srm} summarizes how a version of \gls{scf} that embeds our proposed \gls{srm} works. The algorithm follows the original course of decoding that was described in Section\,\ref{sec:intro_scf}. When the \gls{crc} fails after the initial \gls{sc} pass, the bit estimates and partial sums are stored into restart lists. The bit-flipping candidates are initialized. If during additional trials the bit-flipping index at the \gls{rhs} of the tree identified, the \gls{srm} flag is raised, stored lists are copied to active lists of $\bm{\beta}$ and $\bm{\hat{u}}$. The \gls{sc} decoding is then resumed from the bit $\hat{u}_{\nicefrac{N}{2}}$\,.

The proposed \gls{srm} for \gls{dscf}-$\omega$ decoding with $\omega=1$ follows the same decoding schedule as in Algorithm\,\ref{alg:scf_srm} except the use of the metric function that results on different bit-flipping list. For each bit-flipping set $\bm{\varepsilon}_t = \{i_{\lambda}\}$, only the location of the first bit $i_1$ defines the activation condition of the \gls{srm}. Recall that the bit-flipping indices are added to any set $\bm{\varepsilon}_{t}$ progressively such that $i_1< \ldots < i_{\lambda}$. Therefore, if $i_1$ belongs to the second half of the codeword, the remaining bits of the set are situated there as well.

\gls{srm} can be integrated into other variations of \gls{scf}-based decoders, e.g., those of \cite{fast_sscf,part_scf}. Adapting the mechanism to these decoders would require minimum effort, and it can only  improve their characteristics. This work focuses on \gls{scf} and \gls{dscf}-$\omega$ decoders to demonstrate the functionality. 

\subsection{Memory Structure}
\label{sec:srm_mem}
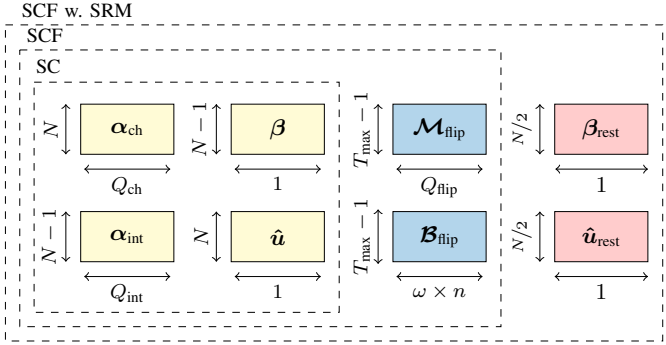
\begin{figure}[t]
\centering
\small
\resizebox{\columnwidth}{!}{\begin{tikzpicture}[]

%% SC
\node [draw, fill=yellow!20, rectangle, align=center, inner sep=3pt,
minimum height=0.7cm, minimum width=1.3cm] (llr_ch) at (0.3,0) {$\bm{\alpha}_{\text{ch}}$}; 

\node [draw, fill=yellow!20, rectangle, align=center, inner sep=3pt,
minimum height=0.7cm, minimum width=1.3cm] (llr_int) at (0.3,-1.5) {$\bm{\alpha}_{\text{int}}$}; 

\node [draw, rectangle, fill=yellow!20, align=center, inner sep=3pt, minimum height=0.7cm, minimum width=1.3cm] (ps) at (2.4,0.0) {$\bm{\beta}$}; 

\node [draw, rectangle, fill=yellow!20, align=center, inner sep=3pt, minimum height=0.7cm, minimum width=1.3cm] (uhat) at (2.4,-1.5) {$\bm{\hat{u}}$}; 

\node [draw, fill=MyBlue!30, rectangle, align=center, inner sep=3pt, minimum height=0.7cm, minimum width=1.3cm] (metr_flip) at ($(ps.east)+(1.6,0)$) {$\bm{\mathcal{M}}_{\text{flip}}$}; 
\node [draw, fill=MyBlue!30, rectangle, align=center, inner sep=3pt, minimum height=0.7cm, minimum width=1.3cm] (bit_flip) at ($(uhat.east)+(1.6,0)$) {$\bm{\mathcal{B}}_{\text{flip}}$};

\node [draw, fill=red!20, rectangle, align=center, inner sep=3pt, minimum height=0.7cm, minimum width=1.3cm] (ps_rest) at ($(metr_flip.east)+(1.6,0)$) {$\bm{\beta}_{\text{rest}}$}; 
\node [draw, fill=red!20, rectangle, align=center, inner sep=3pt, minimum height=0.7cm, minimum width=1.3cm] (uhat_rest) at ($(bit_flip.east)+(1.6,0)$) {$\bm{\hat{u}}_{\text{rest}}$}; 

%% Memory lengths
%%% SC
\draw [<->] ($(llr_ch.south)+(-0.6,-0.2)$) -- node[below,font=\footnotesize]{$Q_{\text{ch}}$} ($(llr_ch.south)+(0.6,-0.2)$);
\draw [<->] ($(llr_ch.west)+(-0.2,-0.35)$) -- node[above,rotate=90,font=\footnotesize]{$N$}($(llr_ch.west)+(-0.2,0.35)$);

\draw [<->] ($(llr_int.west)+(-0.2,-0.35)$) -- node[above,rotate=90,font=\footnotesize]{$N-1$}($(llr_int.west)+(-0.2,0.35)$);
\draw [<->] ($(llr_int.south)+(-0.6,-0.2)$) -- node[below,font=\footnotesize]{$Q_{\text{int}}$} ($(llr_int.south)+(0.6,-0.2)$);

\draw [<->] ($(ps.south)+(-0.6,-0.2)$) -- node[below,font=\footnotesize]{$1$} ($(ps.south)+(0.6,-0.2)$);
\draw [<->] ($(ps.west)+(-0.2,-0.35)$) -- node[above,rotate=90,font=\footnotesize]{$N-1$}($(ps.west)+(-0.2,0.35)$);

\draw [<->] ($(uhat.south)+(-0.6,-0.2)$) -- node[below,font=\footnotesize]{$1$} ($(uhat.south)+(0.6,-0.2)$);
\draw [<->] ($(uhat.west)+(-0.2,-0.35)$) -- node[above,rotate=90,font=\footnotesize]{$N$}($(uhat.west)+(-0.2,0.35)$);

%%% SCF
\draw [<->] ($(metr_flip.south)+(-0.6,-0.2)$) -- node[below,font=\footnotesize]{$Q_{\text{flip}}$}($(metr_flip.south)+(0.6,-0.2)$);
\draw [<->] ($(metr_flip.west)+(-0.2,-0.35)$) -- node[above,rotate=90,font=\footnotesize]{$T_{\text{max}}-1$} ($(metr_flip.west)+(-0.2,0.35)$);

\draw [<->] ($(bit_flip.south)+(-0.6,-0.2)$) -- node[below,font=\footnotesize]{$\omega \times n$} ($(bit_flip.south)+(0.6,-0.2)$);
\draw [<->] ($(bit_flip.west)+(-0.2,-0.38)$) -- node[above,rotate=90,font=\footnotesize]{$T_{\text{max}}-1$}($(bit_flip.west)+(-0.2,0.38)$);

%%% SCF SRM
\draw [<->] ($(ps_rest.south)+(-0.64,-0.2)$) -- node[below]{$1$} ($(ps_rest.south)+(0.65,-0.2)$);
\draw [<->] ($(ps_rest.west)+(-0.2,-0.35)$) -- node[above,rotate=90,font=\footnotesize]{$\nicefrac{N}{2}$}($(ps_rest.west)+(-0.2,0.35)$);

\draw [<->] ($(uhat_rest.south)+(-0.65,-0.2)$) -- node[below]{$1$} ($(uhat_rest.south)+(0.65,-0.2)$);
\draw [<->] ($(uhat_rest.west)+(-0.2,-0.35)$) -- node[above,rotate=90,font=\footnotesize]{$\nicefrac{N}{2}$}($(uhat_rest.west)+(-0.2,0.35)$);

%% Dashed borders
%% SC
\node [font=\footnotesize] (sc_lab) at ($(llr_ch.north) + (-1.1,0.52)$) {\gls{sc}}; 

\draw[dashed] ($(llr_ch.north) + (-1.3,0.3)$) -| ($(uhat.south east)+(0.2,-0.7)$) -| ($(llr_ch.north) + (-1.3,0.3)$);

%SCF
\node [font=\footnotesize] (scf_lab) at ($(llr_ch.north) + (-1.15,0.95)$) {\gls{scf}};

\draw[dashed] ($(llr_ch.north) + (-1.5,0.75)$) -| ($(bit_flip.south east)+(0.2,-0.9)$) -|  ($(llr_ch.north) + (-1.5,0.75)$);

% SCF-SRM
\node [font=\footnotesize] (scf_srm_lab) at ($(llr_ch.north) + (-0.7,1.3)$) {\gls{scf} w. \gls{srm}}; 

%\draw[dashed] ($(llr_ch.north) + (-1.9,1.1)$) -- ($(llr_ch.north)+(5.8,1.1)$) -- ($(llr_ch.north)+(5.8,-4.8)$) -- ($(metr_flip.south) + (-1.9,-0.9)$) -- ($(llr_ch.north) + (-1.9,1.1)$);
\draw[dashed] ($(llr_ch.north) + (-1.7,1.1)$) -| ($(uhat_rest.south east)+(0.2,-1.1)$) -| ($(llr_ch.north) + (-1.7,1.1)$);

\end{tikzpicture}}
\caption{Memory architecture of \gls{scf} embedding the \gls{srm}.}
\label{fig:mem_scf_srm}
\end{figure}

\autoref{fig:mem_scf_srm} shows a memory architecture inspired by \cite{semi_par_sc} and \cite{fast_sscl} for an \gls{scf} decoder that integrates the proposed \gls{srm}. The label indicates the content of the memory. The depth indicates the length of the data vector while the width indicates the number of bits of each entry.
As depicted in \autoref{fig:mem_scf_srm}, \gls{llr}-values and bit-flipping metrics have different quantizations. Channel \glspl{llr} use $Q_{\text{ch}}$ bits, intermediate \glspl{llr}  use $Q_{\text{int}}$ bits and bit-flipping metrics use $Q_{\text{flip}}$ bits. 
The list of bit-flipping candidates requires $ \left(T_{\text{max}}-1\right)\times \omega \times n$ bits with $n$ being the length of the binary representation of a bit-flipping index. The remaining memory blocks are the binary vectors of single bit widths. In total, $\bm{\alpha}_{\text{ch}}$ requires $N\times Q_{\text{ch}}$ bits,  $\bm{\alpha_{\text{int}}}$ requires $\left(N-1\right)\times Q_{\text{int}}$ and $\bm{\mathcal{M}}_{\text{flip}}$ requires $\left(T_{\text{max}}-1\right)\times Q_{\text{flip}}$ bits.
Recall that the \gls{srm} only requires partial sums and bit estimates from the \gls{lhs} tree computed at the initial \gls{sc} trial. Thus, the memory overhead of the \gls{srm}
is $N$ bits.   

\section{Execution-Time Model with Hardware Constraints}
\label{sec:lat_hw_scf}

In order to estimate the latency that reflects an architectural design, we implement a model with a limited number of processing elements, denoted by $P$. The methodology is based on the architecture of the semi-parallel \gls{sc} decoder~\cite{semi_par_sc}.

The nodes of the \gls{sc} decoding tree (\autoref{fig:pol_tree}) perform the calculations of the functions \eqref{eq:f}--\eqref{eq:g} with a limited number of processing elements in parallel. The latency of a single \gls{llr} calculation is considered to be of one \gls{cc}. The vector of the partial-sums is calculated with function  \eqref{eq:ps_comp} for each node in one \gls{cc}. The approach for partial-sums is valid considering simplicity of bitwise XOR operations.

The latency of a single \gls{sc} pass in \glspl{cc} is denoted by $\mathcal{L}_{\text{SC}}$, and it is given by the following equation~\cite{semi_par_sc}:
\begin{equation}
\begin{split}
\mathcal{L}_{\text{SC}} &= \mathcal{L}_{\alpha}+\mathcal{L}_{\beta}\,, \\
&= \left(2N + \frac{N}{P}\cdot \log_2{\left(\frac{N}{4P}\right)}\right) + \left(N-n-1\right)\,,
\end{split}
\label{eq:lat_sc}
\end{equation}
where $n=\log_2\left(N\right)$,  $\mathcal{L}_{\alpha}$ is the latency of \gls{llr} computations and $\mathcal{L}_{\beta}$ is the latency of calculations of partial-sums.

The execution time of one codeword by the \gls{scf} decoder is the product of the \gls{sc} pass latency and the required number of decoding trials. The required number of trials is denoted by $t_{\text{req}}$ and the total execution time is computed as:
\begin{equation}
l_{\text{SCF}} = t_{\text{req}}\cdot \mathcal{L}_{\text{SC}}\,,
\label{eq:lat_scf}
\end{equation}
where $1\leq t_{\text{req}}\leq T_{\text{max}}$. If $t_{\text{req}}=T_{\text{max}}$, $l_{\text{SCF}}$ indicates the worst-case execution time and thus it is decoding latency.

The execution time being variable, the following characteristics are of interest: the average execution time, the average additional execution time and the execution-time variance. These metrics are obtained experimentally, by simulation. The average execution time $\mathcal{L}_{\text{SCF}}$ is estimated by:
\begin{equation}
\mathcal{L}_{\text{SCF}} = \frac{1}{S}\sum_{s=1}^{S} l_{\text{SCF}}\,,
\label{eq:lat_scf}
\end{equation}
where $S$ is the total number of simulated codewords. The average additional execution time $\mathcal{L}'_{\text{SCF}}$ is estimated by: 
\begin{equation}
\mathcal{L}'_{\text{SCF}} = \frac{1}{S'}\sum_{s=1}^{S'} \left(l_{\text{SCF}} - \mathcal{L}_{\text{SC}}\right)\,,
\label{eq:lat_add_scf}
\end{equation}
where  $S' \leq S$ indicates the number of codewords that required more than a single \gls{sc} pass to decode by \gls{scf} decoding. The execution-time variance $\mathcal{V}_l$ is estimated by: 
\begin{equation}
\mathcal{V}_l = \frac{1}{S-1}\sum_{s=1}^{S} \left(l_{\text{SCF}} - \mathcal{L}_{\text{SCF}}\right)^2 \,.
\label{eq:lat_var_scf}
\end{equation}

For \gls{dscf}-$\omega$ decoders, the latency associated with the updates of the bit-flipping sets is ignored since these operations have a negligible impact on the execution time compared to \gls{sc} decoding. Thus, \eqref{eq:lat_scf}\,--\,\eqref{eq:lat_var_scf} are used for the execution-time characteristics of \gls{scf} and \gls{dscf}-$\omega$ decoders.

\section{Simulation Results}
\label{sec:sim_res}

\subsection{Methodology}
A simulation setup is created to analyze the effects of our proposed mechanism on \gls{scf} and \gls{dscf}-$\omega$ decoding. Random blocks of data were encoded with polar codes of $N=1024$ for three different rates $R\in\{\nicefrac{1}{8},\nicefrac{1}{4},\nicefrac{1}{2}\}$, and of $N=512$ for a rate $R=\nicefrac{1}{8}$. A \gls{crc} of $r=16$ bits with polynomial $z^{16}+z^{15}+z^2+1$ is applied.
The polar codes are constructed for a design $\nicefrac{E_b}{N_0}$ of $1.25$\,dB, $1.25$\,dB and $2.5$\,dB for length $N=1024$ of rates $\nicefrac{1}{8}$, $\nicefrac{1}{4}$, and $\nicefrac{1}{2}$, respectively. Polar code for $N=512$ of rate $\nicefrac{1}{8}$ is constructed for a design $\nicefrac{E_b}{N_0}$ of $1.25$\,dB.
Binary phase-shift keying modulation is used over an \gls{awgn} channel. Simulations were run for a minimum of $S=10^5$ random codewords and until $10^3$ frames in errors were found. 
\Gls{dscf}-$\omega$ decoders with $\omega \in \{1,2,3\}$ are examined.

The number of processing elements is limited to $P=64$ for all simulations. 
The \gls{dscf}-$\omega$ decoders use the hardware-friendly \eqref{eq:metr_dscf_w_simp} function for metric calculations. 
The maximum number of trials $T_{\text{max}}$ is set to $13$ for \gls{scf}, while for \gls{dscf}-$\omega$ they are set to $T_{\text{max}}\in \{8,51,301\}$ for $\omega\in\{1,2,3\}$. The values of $T_{\text{max}}$ were selected to achieve an error-correction performance that is close to the genie-aided decoder~\cite{dyn_scf} at the target \gls{fer}.

We compare the decoders for $\mathcal{P}(1024,128)$ polar code with $r=16$ with and without the \gls{srm} in terms of error-correction performance, execution-time characteristics and memory requirements. 
We highlight the results of the execution-time characteristics of using \gls{srm} for the target \gls{fer} of $10^{-2}$ for polar codes of different code lengths and code rates.

\subsection{Error-Correction Performance}
\begin{figure}[t]
\centering
\begin{tikzpicture}
  \pgfplotsset{
    label style = {font=\fontsize{9pt}{7.2}\selectfont},
    tick label style = {font=\fontsize{9pt}{7.2}\selectfont}
  }

  \begin{semilogyaxis}[%
    width=\columnwidth,
    height=\plotfigureheight\columnwidth,
    xmin=0.25, xmax=2.0,
    xlabel={$\nicefrac{E_b}{N_0},\,\mathrm{dB}$},
    xlabel style={yshift=0.4em},
    ymin=2e-5, ymax=1,
    ylabel style={yshift=-0.1em},
    ylabel={Frame-error rate},
    yminorticks, xmajorgrids,
    ymajorgrids, yminorgrids,
    legend pos=south west,
    legend style={legend columns=2, font=\footnotesize, column sep=0mm, row sep=-0.5mm},
    mark size=1.6pt, mark options=solid,
    ]

\addplot[color=black, mark=diamond, line width=0.8pt, mark size=2.1pt]
    table[row sep=crcr]{%
    0.25  0.533780\\
    0.5   0.401080\\
    0.75  0.276750\\
    1.0   0.178000\\
    1.25  0.102700\\
    1.5   0.054850\\
    1.75  0.026920\\
    2.0   0.011660\\
    2.25  0.004823\\
    2.5   0.001827\\
    };   
    \addlegendentry{\gls{scf}}

    \addplot[color=black, dashed,  mark=x, line width=0.8pt, mark size=2.1pt]
    table[row sep=crcr]{%
    0.25  0.533780\\
    0.5   0.401080\\
    0.75  0.276750\\
    1.0   0.178000\\
    1.25  0.102700\\
    1.5   0.054850\\
    1.75  0.026920\\
    2.0   0.011660\\
    2.25  0.004823\\
    2.5   0.001827\\
    };   
    \addlegendentry{\gls{scf} w. \gls{srm}}

 \addplot[color=red, mark=pentagon, line width=0.8pt, mark size=2.1pt]
    table[row sep=crcr]{%
    0.25  0.544050\\
    0.5   0.398670\\
    0.75  0.268450\\
    1.0   0.163360\\
    1.25  0.087870\\
    1.5   0.042450\\
    1.75  0.019100\\
    2.0   0.007371\\
    2.25  0.002819\\
    2.5   0.001005\\
    2.75  0.000302\\
    };   
    \addlegendentry{\gls{dscf}-1}

    \addplot[color=red, mark=x, dashed, line width=0.8pt, mark size=2.1pt]
    table[row sep=crcr]{%
    0.25  0.544050\\
    0.5   0.398670\\
    0.75  0.268450\\
    1.0   0.163360\\
    1.25  0.087870\\
    1.5   0.042450\\
    1.75  0.019100\\
    2.0   0.007371\\
    2.25  0.002819\\
    2.5   0.001005\\
    2.75  0.000302\\
    };   
    \addlegendentry{\gls{dscf}-1 w. \gls{srm}}

    \addplot[color=MyBlue, mark=triangle, line width=0.8pt, mark size=2.1pt]
    table[row sep=crcr]{%
    0.2500  0.382990\\
    0.5000  0.246460\\
    0.7500  0.140310\\
    1.0000  0.070650\\
    1.2500  0.030570\\
    1.5000  0.011620\\
    1.7500  0.004046\\
    2.0000  0.001174\\
    2.25  0.000329\\
    };   
    \addlegendentry{\gls{dscf}-2} 
    
    \addplot[color=MyBlue, dashed, mark=x, line width=0.8pt, mark size=2.1pt]
    table[row sep=crcr]{%
    0.2500  0.382990\\
    0.5000  0.246460\\
    0.7500  0.140310\\
    1.0000  0.070650\\
    1.2500  0.030570\\
    1.5000  0.011620\\
    1.7500  0.004046\\
    2.0000  0.001174\\
    2.25  0.000329\\
    };   
    \addlegendentry{\gls{dscf}-2 w. \gls{srm}} 

    \addplot[color=MyDarkGreen, mark=square, line width=0.8pt, mark size=2.1pt]
    table[row sep=crcr]{%
    0.2500  0.260600\\
    0.5000  0.146490\\
    0.7500  0.072610\\
    1.0000  0.031110\\
    1.2500  0.011720\\
    1.5000  0.003501\\
    1.7500  0.000986\\
    2.0000  0.000232\\
    };   
    \addlegendentry{\gls{dscf}-3}

    \addplot[color=MyDarkGreen, dashed, mark=x, line width=0.8pt, mark size=2.1pt]
    table[row sep=crcr]{%
    0.2500  0.260600\\
    0.5000  0.146490\\
    0.7500  0.072610\\
    1.0000  0.031110\\
    1.2500  0.011720\\
    1.5000  0.003501\\
    1.7500  0.000986\\
    2.0000  0.000232\\
    };   
    \addlegendentry{\gls{dscf}-3 w. \gls{srm}} 
    
  \end{semilogyaxis}

\end{tikzpicture}%
\caption{Error-correction performance of \gls{scf} and \gls{dscf}-$\omega$ decoders with $\omega \in \{1,2,3\}$ with and without the \gls{srm} for $\mathcal{P}(1024,128)$ polar code with $r=16$.}
\label{fig:fer_scf_dscfw}
\end{figure}
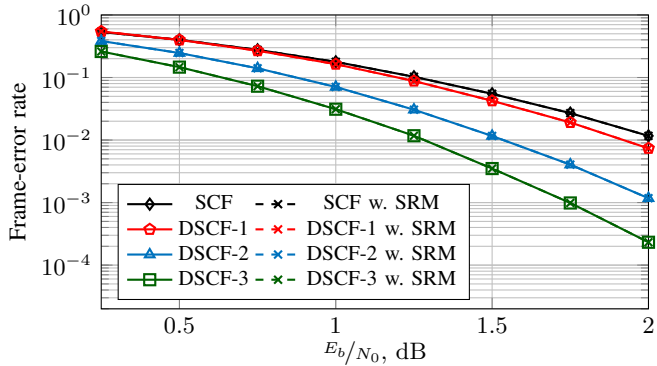

The error-correction performance in terms of \gls{fer} for \gls{scf} and \gls{dscf}-$\omega$ decoders is shown in \autoref{fig:fer_scf_dscfw} for $\mathcal{P}(1024,128)$ polar code. The decoders with and without the \gls{srm} are depicted in dashed and solid lines, respectively. The \gls{scf} decoder is in black, \gls{dscf}-2 is in light-blue and \gls{dscf}-3 is in light-green with unique markers.

From \autoref{fig:fer_scf_dscfw}, it can be seen that the \gls{srm} does not alter the error-correction performance. This is expected and in line with the definition of the mechanism described in Section \ref{subsec:SRM_description}. The results also agree with~\cite{dyn_scf}, i.e., they indicate that \gls{dscf}-$\omega$ outperforms standard \gls{scf}. The \gls{dscf}-3 decoder offers the best performance, thus the motivation to reduce the execution-time characteristics of that algorithm.

\subsection{Execution-Time Characteristics}

\begin{figure}[t]
\centering
\begin{tikzpicture}

  \pgfplotsset{
    label style = {font=\fontsize{9pt}{7.2}\selectfont},
    tick label style = {font=\fontsize{9pt}{7.2}\selectfont}
  }

   \begin{semilogyaxis}[%
    width=\columnwidth,
    height=\plotfigureheight\columnwidth,
    xmin=3.2e-04, xmax=1.05e-01,
    xlabel={Frame-error rate},
    xlabel style={yshift=0.4em},
    ymin=2e3, ymax=2e5,
    x dir=reverse,
    xmode=log,
    ylabel style={yshift=-0.1em},
    ylabel={Avg. Exec. Time, $\mathcal{L}_{\text{SCF}}$},
    xlabel style={yshift=-0.2em},
    yminorticks, xmajorgrids,
    ymajorgrids, yminorgrids,
    legend style={at={(1.0,1.0)},anchor=north east},
    legend style={legend columns=2, font=\scriptsize, column sep=0mm, row sep=-0.5mm}, 
    mark size=1.8pt, mark options=solid,
    ] 
    
 \addplot [line width=0.8pt, mark=diamond, color=black]
 table[x=fer,y=execav]{data/scf/fer_scf_lowr_aver_exec_time_log.tex};
  \addlegendentry{\gls{scf}}
  
  \addplot [dashed, line width=0.8pt, mark=diamond, color=black]
 table[x=fer,y=execavsrm]{data/scf/fer_scf_lowr_aver_exec_time_log.tex};
  \addlegendentry{\gls{scf} w. \gls{srm}}
 
  \addplot [line width=0.8pt, mark=pentagon, color=red]
 table[x=fer,y=execav]{data/dscf1/fer_dscf1_lowr_aver_exec_time_log.tex};
  \addlegendentry{\gls{dscf}-1}
  
  \addplot [dashed, line width=0.8pt, mark=pentagon, color=red]
 table[x=fer,y=execavsrm]{data/dscf1/fer_dscf1_lowr_aver_exec_time_log.tex};
  \addlegendentry{\gls{dscf}-1 w. \gls{srm}}

  \addplot [line width=0.8pt, mark=triangle, color=MyBlue]
 table[x=fer,y=execav]{data/dscf2/fer_dscf2_lowr_aver_exec_time_log.tex};
  \addlegendentry{\gls{dscf}-2}
  
  \addplot [dashed, line width=0.8pt, mark=triangle, color=MyBlue]
 table[x=fer,y=execavsrm]{data/dscf2/fer_dscf2_lowr_aver_exec_time_log.tex};
  \addlegendentry{\gls{dscf}-2 w. \gls{srm}}
  
    \addplot [line width=0.8pt, mark=square, color=MyDarkGreen]
  table[x=fer,y=execav]{data/dscf3/fer_dscf3_lowr_aver_exec_time_log.tex};
   \addlegendentry{\gls{dscf}-3}
  
  \addplot [dashed, line width=0.8pt, mark=square, color=MyDarkGreen]
 table[x=fer,y=execavsrm]{data/dscf3/fer_dscf3_lowr_aver_exec_time_log.tex};
  \addlegendentry{\gls{dscf}-3 w. \gls{srm}}
    
    \addplot[color=MyOrange,  mark=none, line width=0.8pt] table[row sep=crcr]{%
    1 3157\\    
    1e-1 3157\\    
    1e-2 3157\\    
    1e-3 3157\\    
    3e-4 3157\\    
    };
    \addlegendentry{\gls{sc}}

\end{semilogyaxis}    
\end{tikzpicture}%
\caption{Average execution time of \gls{scf} and  \gls{dscf}-$\omega$ decoders with $\omega\in\{1,2,3\}$ with and without the \gls{srm} for $\mathcal{P}(1024,128)$ polar code with $r=16$. Plain \gls{sc} decoder is for reference.}
\label{fig:lat_tot_scf_dscf_srm_log_lowrate}
\end{figure}
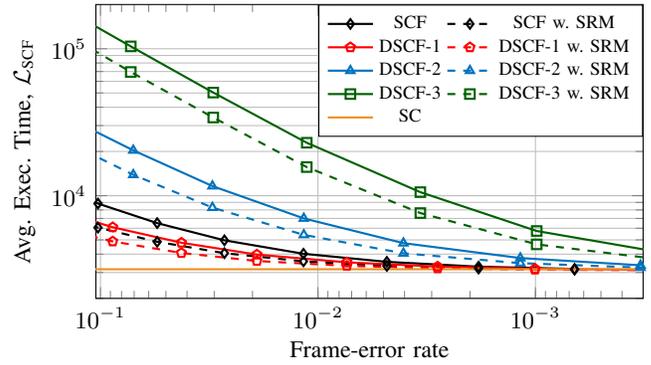

\autoref{fig:lat_tot_scf_dscf_srm_log_lowrate} show the average execution time 
of \gls{scf} and \gls{dscf}-$\omega$ decoders for $\mathcal{P}(1024,128)$ polar code. The decoders with and without the \gls{srm} are in dashed and solid lines accordingly. %The colors and the marks  of \autoref{fig:fer_scf_dscfw} remain.
The latency of \gls{sc} decoding is provided for reference. 
From the figures, we observe that using the \gls{srm} provides greater gain to the \gls{dscf}-$2$ and \gls{dscf}-$3$ decoders throughout the \gls{fer} range. We explain this by higher number of additional trials performed by the multi bit-flipping decoders in average. We also observe that \gls{dscf}-1 decoder with \gls{srm} provides the smallest reduction among the other decoders. This can be explained by low number of additional decoding attempts and low decoding latency (smallest $T_{\text{max}}$ among the other decoders). 
In \autoref{fig:lat_tot_scf_dscf_srm_log_lowrate} we can also see that at lower \gls{fer} the average execution time of all decoders with and without the  \gls{srm} closely approach the latency of  the \gls{sc} decoder.

The reduction of the execution-time characteristics are summarized in \autoref{tab:gain_scf_all} for each decoder for polar codes of $N=1024$ for various code rates. \autoref{tab:gain_scf_all_N512} shows results of $N=512$ for $R=\nicefrac{1}{8}$ exhibiting the higher gain with the \gls{srm}. The notation for the characteristics is described in Section\,\ref{sec:lat_hw_scf}.
The differences are denoted by $\Delta$ and presented in percent. 
The $\nicefrac{E_b}{N_0}$ points for each decoder are indicated that correspond to target \gls{fer} of $10^{-2}$. 
The results provided in tables indicate the greatest gain from applying the \gls{srm} for polar code of $N=1024$ for $R=\nicefrac{1}{8}$. Compared to the original \gls{scf} decoder, for the \gls{scf} embedding the \gls{srm}, the average execution time, the average additional execution time and the execution-time variance are reduced by $11.17\%$, $48.30\%$, and $73.50\%$, respectively. \gls{dscf}-1 decoder embedding \gls{srm} provides reduction of $7.57\%$, $43.57\%$ and $67.28\%$, respectively. 
Highlighting the results for \gls{dscf}-3 -- decoder with the strongest error-correction performance, applying the \gls{srm} provides reduction of $31.70\%$, $37.08\%$ and $57.28\%$, respectively. Looking at results for polar codes of $N=1024$ for higher code rates, we see that, while the reductions of the execution-time characteristics are lower, the general tendencies are preserved. Comparing results for polar codes of $N=1024$ and $N=512$  for $R=\nicefrac{1}{8}$, the reduction of characteristics is almost identical. Applying the \gls{srm} to \gls{dscf}-3 decoder for polar code of $N=1024$ for $R=\nicefrac{1}{2}$ provides the reduction of $7.33\%$, $9.03\%$ and $12.20\%$, respectively.

\subsection{Memory Estimates}
The memory is calculated as described in Section\,\ref{sec:srm_mem}, where the same quantization scheme as that of \cite{pract_dscf} is used. Hence, channel \glspl{llr} and intermediate \glspl{llr}
and bit-flipping metrics are quantized by $Q_{\text{ch}}=6$, $Q_{\text{int}}=7$ and $Q_{\text{flip}}=7$ bits, respectively. Out of these bits, $2$ bits of $Q_{\text{ch}}$ and $Q_{\text{int}}$ are used for the fractional part while $3$ bit are used for $Q_{\text{flip}}$.  The memory estimates and memory overhead in percent are provided in \autoref{tab:mem_scf_srm} for all considered decoders. The results are provided for polar codes of different lengths, while the code rate does not affect the memory size. It can be seen from \autoref{tab:mem_scf_srm} that the proposed \gls{srm} leads to a memory overhead of $2.86\%$ to $6.62\%$.  
Embedding the \gls{srm} into \gls{dscf}-3 decoder results in the smallest memory overhead compared to the other decoders, since \gls{dscf}-3 requires a much larger memory to store the list of bit-flipping candidates $\bm{\mathcal{B}}_{\text{flip}}$ and the  corresponding list of bit-flipping metrics $\bm{\mathcal{M}}_{\text{flip}}$\,.
\begin{table*}[t]
\centering
\caption{Reduction of the execution-time characteristics by using \gls{srm} to \gls{scf} and \gls{dscf}-$\omega$ decoders for polar codes of $N=1024$ for various code rates at the target \gls{fer} $10^{-2}$.}
\vspace{-1.5mm}
\setlength{\tabcolsep}{5pt} % Default value: 6pt
\renewcommand{\arraystretch}{1.4} % Default value: 1
\begin{tabular}{|c|c|c|c|c|c||c|c|c|c||c|c|c|c|}
\cline{3-14}
\multicolumn{2}{c}{}&\multicolumn{4}{|c||}{$\mathcal{P}\left(1024,128\right)$, $r=16$} & \multicolumn{4}{c||}{$\mathcal{P}\left(1024,256\right)$, $r=16$} & \multicolumn{4}{c|}{$\mathcal{P}\left(1024,512\right)$, $r=16$} \\ \cline{2-14}
\multicolumn{1}{c|}{} & $T_{\text{max}}$ &  $\nicefrac{E_b}{N_0}$ & $\Delta \mathcal{L}_{\text{SCF}}$ & $\Delta \mathcal{L}'_{\text{SCF}}$ & $\Delta \mathcal{V}_{l}$ & $\nicefrac{E_b}{N_0}$ & $\Delta \mathcal{L}_{\text{SCF}}$ & $\Delta \mathcal{L}'_{\text{SCF}}$ & $\Delta \mathcal{V}_{l}$  & $\nicefrac{E_b}{N_0}$ & $\Delta \mathcal{L}_{\text{SCF}}$ & $\Delta \mathcal{L}'_{\text{SCF}}$ & $\Delta \mathcal{V}_{l}$  \\ \hline
\cellcolor{black!50}\gls{scf} & $13$ & $2.0$ & $11.17$ & $48.30$ & $73.50$ & $1.875$ & $9.00$ & $44.68$ & $69.15$ & $2.375$& $6.54$& $37.44$ & $59.71$\\ \hline
\cellcolor{red!50}\gls{dscf}-1 & $8$ & $1.875$ & $7.57$ & $43.57$  & $67.28$ & $1.75$ & $5.01$ & $33.33$ & $50.87$ & $2.25$ & $2.95$ & $22.04$ &$32.35$\\ \hline
\cellcolor{MyBlue!50}\gls{dscf}-2 & $51$ & $1.5$ & $22.73$ & $40.71$ & $63.26$ & $1.44$ & $14.74$ & $27.60$ & $41.74$ & $2.00$& $6.96$ & $13.83$ & $19.00$\\ \hline
\cellcolor{MyDarkGreen!50}\gls{dscf}-3& $301$ & $1.25$ & $31.70$ & $37.08$ & $57.28$ & $1.25$ &  $19.09$ &  $22.57$  & $34.78$ & $1.875$ & $7.33$& $9.03$&$12.20$\\ \hline 
\multicolumn{2}{c|}{} & in $\mathrm{dB}$ & \multicolumn{3}{c||}{$\Delta$ in $\%$} & \multicolumn{1}{c}{in $\mathrm{dB}$} & \multicolumn{3}{|c||}{$\Delta$ in $\%$} & \multicolumn{1}{c}{in $\mathrm{dB}$} & \multicolumn{3}{|c|}{$\Delta$ in $\%$} \\
\cline{3-14}%\cline{4-6} \cline{8-10}\cline{12-14}
\end{tabular}
\label{tab:gain_scf_all}
\end{table*}

\begin{table}[h]
\centering
\caption{Reduction of the execution-time characteristics by using \gls{srm} to \gls{scf} and \gls{dscf}-$\omega$ decoders for polar codes of $N=512$ for $R=\nicefrac{1}{8}$ at the target \gls{fer} $10^{-2}$.}
\vspace{-1.5mm}
\setlength{\tabcolsep}{5pt} % Default value: 6pt
\renewcommand{\arraystretch}{1.4} % Default value: 1
\begin{tabular}{|c|c|c|c|c|c|}
\cline{3-6}
\multicolumn{2}{c}{}&\multicolumn{4}{|c|}{$\mathcal{P}\left(512,64\right)$, $r=16$} \\ \cline{2-6}
\multicolumn{1}{c|}{} & $T_{\text{max}}$ &  $\nicefrac{E_b}{N_0}$ & $\Delta \mathcal{L}_{\text{SCF}}$ & $\Delta \mathcal{L}'_{\text{SCF}}$ & $\Delta \mathcal{V}_{l}$ \\ \hline
\cellcolor{black!50}\gls{scf} & $13$ & $2.625$ & $10.36$ & $47.49$ & $72.27$ \\ \hline
\cellcolor{red!50}\gls{dscf}-1 & $8$ & $2.625$ & $6.03$ & $43.93$ & $67.94$ \\ \hline
\cellcolor{MyBlue!50}\gls{dscf}-2 & $51$ & $2.0$ & $24.01$ & $40.52$ & $63.08$ \\ \hline
\cellcolor{MyDarkGreen!50}\gls{dscf}-3& $301$ & $1.75$ & $32.33$ & $37.59$ & $58.67$ \\ \hline 
\multicolumn{2}{c|}{} & in $\mathrm{dB}$ & \multicolumn{3}{c|}{$\Delta$ in $\%$}  \\
\cline{3-6}
\end{tabular}
\label{tab:gain_scf_all_N512}
\end{table}

\glsreset{srm}
\section{Conclusion}
\label{sec:conclusion}
In this work, we proposed \gls{srm}, a mechanism that reduces the execution-time characteristics of \gls{scf} and \gls{dscf}-$\omega$ decoders by starting trials beyond the initial one from the middle of the decoding process if the flipping index falls into the \glsentrylong{rhs} of the decoding tree. 
The mechanism requires to store a small amount of results from the initial \gls{sc} pass after the \glsentrylong{lhs} of the tree has been visited. We showed the minor modifications required to use it in a \gls{dscf}-$\omega$ decoder. The proposed mechanism can be integrated to other \gls{scf}-based decoding algorithms, does not affect the error-correction performance, and works with any code length and rate. For a \gls{dscf}-$3$ decoder for polar code of length $1024$ bits, the average execution time, the average additional execution time and the execution-time variance were shown to be reduced by $31\%$, $37\%$ and $57\%$, respectively, at the cost of a $3.9\%$ memory overhead.

\section*{Acknowledgement}
The authors want to thank Tannaz Kalatian for her initial work on the topic.
Work supported by NSERC Discovery Grant \#651824.

\bibliographystyle{IEEEtran}
%% Use the standard abreviations as much as possible.
\bibliography{IEEEabrv,ConfAbrv,refs}
\begin{table}[t]
\centering
\caption{Memory estimates and memory overhead of decoders caused by \gls{srm} for polar codes of $N\in\{1024,512\}$.}
\vspace{-1.5mm}
\setlength{\tabcolsep}{6pt} % Default value: 6pt
\renewcommand{\arraystretch}{1.3} % Default value: 1
\begin{tabular}{|c|c|c|c|c|}
\multicolumn{5}{c}{Polar code with $N=1024$} \\ \cline{2-5}
\multicolumn{1}{c|}{} & $T_{\text{max}}$ & no \gls{srm},\,bits & w. \gls{srm},\,bits & mem. incr.,\,\% \\ \hline
\cellcolor{black!50}\gls{scf} & $13$ & $15556$  & $16580$  & $6.58$  \\ \hline
\cellcolor{red!50}\gls{dscf}-1 & $8$ & $15471$  & $16495$  & $6.62$  \\ \hline 
\cellcolor{MyBlue!50}\gls{dscf}-2 & $51$ & $16702$ & $17726$ & $6.13$   \\ \hline 
\cellcolor{MyDarkGreen!50}\gls{dscf}-3 & $301$ & $26452$ & $27476$ & $3.87$  \\ \hline 
\multicolumn{5}{c}{Polar code with $N=512$} \\ \cline{2-5}
\multicolumn{1}{c|}{} & $T_{\text{max}}$ & no \gls{srm},\,bits & w. \gls{srm},\,bits & mem. incr.,\,\% \\ \hline
\cellcolor{black!50}\gls{scf} & $13$ &  $7864$ & $8376$  & $6.51$   \\ \hline
\cellcolor{red!50}\gls{dscf}-1 & $8$ & $7784$  & $8296$  & $6.58$  \\ \hline 
\cellcolor{MyBlue!50}\gls{dscf}-2 & $51$ & $8922$ & $9434$ & $5.74$  \\ \hline 
\cellcolor{MyDarkGreen!50}\gls{dscf}-3 & $301$ & $17872$ & $18384$ & $2.86$  \\ \hline 
\end{tabular}
\label{tab:mem_scf_srm}
\end{table}

\balance % Balance the columns on the last page. Requires balance package.
\end{document}